\documentclass[journal,twoside,web]{ieeecolor}

\usepackage{generic}
\usepackage{cite}
\usepackage{amsmath,amssymb,amsfonts}

\usepackage{graphicx}
\usepackage{multirow}
\usepackage{array} 
\usepackage[linesnumbered,ruled,vlined]{algorithm2e}

\usepackage{hyperref}
\hypersetup{hidelinks=true}
\usepackage{orcidlink}

\def\BibTeX{{\rm B\kern-.05em{\sc i\kern-.025em b}\kern-.08em
    T\kern-.1667em\lower.7ex\hbox{E}\kern-.125emX}}
\begin{document}
\SetKwInput{KwInput}{Input}

\title{Improving Deep Learning–based Respiratory Sound Analysis with Frequency Selection and Attention Mechanism}
\author{Nouhaila Fraihi\,\orcidlink{0009-0007-2815-0448}, 
        Ouassim Karrakchou\,\orcidlink{0000-0002-9082-0736},~\IEEEmembership{Senior Member, IEEE}, 
        and Mounir Ghogho\,\orcidlink{0000-0002-0055-7867},~\IEEEmembership{Fellow, IEEE}%
\thanks{This research was partially funded by Mohammed VI Polytechnic University (UM6P) through the i-Respire research project.}
\thanks{Nouhaila Fraihi and Ouassim Karrakchou are with the TICLab, International University of Rabat, Morocco (emails: nouhaila.fraihi@uir.ac.ma; ouassim.karrakchou@uir.ac.ma).}
\thanks{Mounir Ghogho is with the College of Computing, Mohammed VI Polytechnic University (UM6P), Morocco, and also with the Faculty of Engineering, University of Leeds, UK (emails: mounir.ghogho@um6p.ma; m.ghogho@leeds.ac.uk).}
\thanks{An earlier version of this work was published in the EMBC conference: N. Fraihi,O.Karrakchou ,M.Ghogho, ``Enhancing Efficiency in CNN-Based Respiratory Sound Analysis through Temporal Self-Attention and Frequency Band Selection,'' \textit{Proc.\ IEEE EMBC}, 2025}
}

\maketitle


\begin{abstract}
Accurate classification of respiratory sounds requires deep learning models that effectively capture fine-grained acoustic features and long-range temporal dependencies. Convolutional Neural Networks (CNNs) are well-suited for extracting local time–frequency patterns but are limited in modeling global context. In contrast, transformer-based models can capture long-range dependencies, albeit with higher computational demands. To address these limitations, we propose a compact CNN-Temporal Self-Attention (CNN-TSA) network that integrates lightweight self-attention into an efficient CNN backbone. Central to our approach is a Frequency Band Selection (FBS) module that suppresses noisy and non-informative frequency regions, substantially improving accuracy and reducing FLOPs by up to 50\%. We also introduce age-specific models to enhance robustness across diverse patient groups. Evaluated on the SPRSound‑2022/2023 and ICBHI‑2017 lung sound datasets, CNN–TSA with FBS sets new benchmarks on SPRSound and achieves state-of-the-art performance on ICBHI, all with a significantly smaller computational footprint. Furthermore, integrating FBS into an existing transformer baseline yields a new record on ICBHI, confirming FBS as an effective drop-in enhancement. These results demonstrate that our framework enables reliable, real-time respiratory sound analysis suitable for deployment in resource-constrained settings.

\end{abstract} 


\section{Introduction}

\IEEEPARstart{R}{espiratory} diseases remain a leading source of global morbidity and mortality, highlighting the demand for precise diagnostic tools \cite{momtazmanesh2023global}. Lung-sound analysis plays a crucial role in assessing pulmonary function, as respiratory acoustics reflect pulmonary status \cite{pasterkamp1997respiratory}; yet, conventional auscultation is constrained by the clinician’s subjective interpretation \cite{sarkar2015auscultation}. The advent of digital stethoscopes and AI has, therefore, shifted diagnostics toward automated, objective interpretation \cite{siebert2023deep}.

Traditional machine learning approaches have been instrumental in classifying adventitious respiratory sounds but often struggle to capture their complex acoustic features \cite{garcia2023machine}. Deep learning (DL) architectures have emerged as a powerful alternative, demonstrating superior performance in extracting and modeling the intricate patterns essential for accurate analysis of respiratory sounds \cite{sabry2024lung}.

One prevalent approach involves using Convolutional Neural Networks (CNNs) to analyze the time-frequency representations \cite{allen1977short} of audio signals \cite{moummad2023pretraining,nguyen2022lung,li2022explainable,purwins2019deep,chen2022classify}. A major advantage of CNNs in conventional image classification is their effectiveness at capturing local dependencies \cite{lecun2015deep}. However, audio spectrograms demonstrate significant long-range harmonic dependencies along both frequency and time dimensions\cite{choi2017tutorial}. Specifically, respiratory sounds feature cyclic patterns involving long-range temporal dependencies, essential for distinguishing between normal and adventitious sounds \cite{marques2018normal}.

To handle those properties of respiratory spectrograms, transformer-based models have been introduced \cite{mang2024classification,bae2023patch,he2024multi}. Models such as the Audio Spectrogram Transformer (AST) \cite{gong2021ast} partition spectrograms into discrete patches and employ self‑attention to model global temporal–spectral interactions.However, despite their representational power, they incur substantial computational overhead, which hampers real‑time inference and constrains deployment on resource‑limited hardware typical of many clinical settings \cite{khan2022transformers}.

In this study, we aim to leverage the advantages of both CNN and transformer techniques while addressing their shortcomings in the context of respiratory sounds classification. Hence, we propose a hybrid architecture that combines CNNs with self-attention mechanisms for effective analysis of respiratory sound spectrograms. CNNs are employed to extract fine-grained local features in the time-frequency domain, while self-attention addresses their limitation in modeling long-range temporal dependencies. 

Another major challenge in respiratory sound analysis is the presence of noise and overlapping events across a wide frequency range, including coughing, speech, heartbeats, alarms, and device artifacts~\cite{emmanouilidou2017computerized,kim2021respiratory}. These disruptions hinder accurate event prediction~\cite{kochetov2018noise}. To mitigate their impact, it is crucial to isolate and leverage only the most relevant features, ensuring that non-essential signals do not degrade model performance. This underscores the need for feature selection methods that enhance both accuracy and generalization.

While many methods exist for evaluating feature importance in CNNs~\cite{yang2023survey}, most are designed for explainable AI in natural image tasks~\cite{zhang2020measuring} and serve primarily as qualitative visualization tools. Although some selection techniques based on importance scores have been proposed~\cite{sosa2024feature}, none, to the best of our knowledge, specifically target frequency selection for respiratory sound spectrograms.

To address this gap, we propose a frequency band selection method tailored to the spectral properties of respiratory spectrograms. By quantifying each band's contribution to model predictions, irrelevant frequencies can be excluded, enhancing generalization and reducing computational load while preserving the most informative spectral features.

Moreover, respiratory sounds vary across age groups due to differences in airway structure and lung function; children tend to produce higher-pitched, less stable sounds, while adults exhibit lower, more consistent patterns~\cite{gross2000relationship,marques2018normal}. To account for these variations, we propose age-specific models for pediatric and adult populations. Given that age is readily available, this approach is both practical and effective for improving classification accuracy.


The contributions of our research are summarized as follows:

\begin{enumerate}
\itemsep=0pt
\item We propose a CNN Temporal Self-Attention (CNN-TSA) network for respiratory event classification, which combines CNNs with temporal self-attention to capture both local features and long-range temporal dependencies in Mel spectrograms.

\item We introduce an importance-based frequency band selection (FBS) method that improves generalization and reduces computation, outperforming standard backward selection in both speed and accuracy.

\item We develop age-specific models to account for acoustic variations across age groups.

\item Our framework achieves state-of-the-art results on the SPRSound dataset and competitive performance on ICBHI, using a compact architecture, and the FBS reduces FLOPs by up to 50\%.

\item We show that our FBS method applies to other architectures. Most notably, we applied FBS in a transformer-based setting (Patch-Mix + AST~\cite{bae2023patch}) and achieved new benchmarks on ICBHI.
\end{enumerate}

The structure of this paper is as follows: Section \ref{sec:related works} reviews related work on deep learning-based respiratory sound analysis. Section \ref{sec:method} presents and details the proposed methods. Section \ref{sec:experimental settings} describes the experimental setup, including datasets, evaluation metrics, and preprocessing steps. Section \ref{sec:experimental results} reports the results, including an ablation study and a comparison with state-of-the-art approaches. Finally, Sections \ref{sec:discussion} and \ref{sec:conclusion} discuss the findings and conclude the article.

\section{Related works}

\label{sec:related works}

In this study, we review recent advancements in respiratory sound analysis using deep learning, focusing on CNN- and transformer-based models. We also highlight feature importance-based feature selection.

\subsection{Deep Learning-Based Respiratory Sound Analysis}

The development of large, annotated datasets has been instrumental in this progress. In particular, the ICBHI database \cite{rocha2019open} from the ICBHI 2017 challenge emphasizes performance comparison, while the SPRSound database \cite{zhang2022sprsound}, which is part of the IEEE BioCAS challenges in 2022 \cite{zhang2022grand} and 2023 \cite{zhang2022sprsound}, seeks to improve both performance and real-time inference efficiency. These challenges underscore the importance of model generalization to ensure consistent performance across varied patient populations.

Several studies have investigated CNN-based methods for respiratory sound analysis. For instance, Supervised Contrastive Learning (SCL) has been employed to optimize embeddings by incorporating demographic information \cite{moummad2023pretraining}. To address dataset limitations, variants of ResNet-34 with device-specific fine-tuning and data augmentation strategies have been proposed \cite{gairola2021respirenet}. Lightweight CNNs using Mel-Frequency Cepstral Coefficients (MFCCs) have enabled efficient audio processing \cite{babu2022multiclass}, while pre-trained ResNet models enhanced with co-tuning and stochastic normalization have demonstrated improved classification performance \cite{nguyen2022lung}. Furthermore, CNNs integrated within Mixture of Experts (MoE) frameworks have shown effectiveness in enhancing respiratory cycle recognition \cite{pham2021cnn}, and the combination of Short-Time Fourier Transform (STFT) features with fine-tuned ResNet18 models has proven beneficial for detecting abnormal lung sounds \cite{chen2022classify}. Despite these advances, CNN-based methods often fail to account for the fundamental differences between spectrograms and natural images, particularly the unique semantic structure and the distinct temporal and frequency axes inherent in spectrogram representations.To overcome the limitations of single-domain CNN models, dual-input architectures have been explored, leveraging both time-domain and frequency-domain features. One such approach combines raw audio signals processed with a 1D CNN and STFT spectrograms analyzed by a 2D CNN \cite{pessoa2023pediatric}. Another method employs a dual-route design, where a ResNet-50 extracts features from log Mel-spectrograms while a parallel 1D CNN captures temporal dynamics from raw audio data \cite{taghibeyglou2024trespnet}. Beyond architectural enhancements and multi-domain feature integration, attention mechanisms have also been incorporated into CNN-based frameworks to further enhance respiratory sound classification. For example, one study integrates Squeeze-and-Excitation (SE), spatial attention, their combination, and component attention blocks within a ResNet-18 architecture \cite{yang2020adventitious}. Another work improves performance by embedding an Efficient Channel Attention (ECA-Net) module into a modified VGGish model utilizing log-Mel spectrogram and MFCC features \cite{choi2023interpretation}. Additionally, a system combining Continuous Wavelet Transformation with Inception-residual-based models leverages spatiotemporal focusing and multi-head attention to improve feature representation \cite{ngo2023deep}. Despite these advancements, many CNN-based methods continue to overlook key structural differences between spectrograms and natural images, particularly the long-range temporal dependencies inherent in respiratory audio.

To address these limitations, transformer-based architectures have been explored, offering a more suitable framework for capturing such dependencies by treating spectrograms as sequences of patches and applying self-attention for feature extraction. One notable approach adapts the Vision Transformer (ViT) framework with variably sized patches to better capture the spectral and temporal characteristics of audio signals \cite{he2024multi}. Another employs Patch-Mix augmentation with a pre-trained AST model to enhance robustness and feature diversity \cite{bae2023patch}. The use of cochleograms as input to ViTs has also been proposed to exploit richer spectro-temporal features \cite{mang2024classification}. Furthermore, a self-supervised method, Masked Modeling Duo (M2D), improves representation learning by separately encoding masked regions of the input using a ViT-based architecture \cite{niizumi2024masked}. In addition, a multimodal approach within the CLAP framework incorporates patient demographics and recording conditions alongside audio data to enable context-aware classification \cite{kim2024bts}. However, the substantial computational demands of transformer-based models present challenges for real-time inference and deployment in resource-constrained clinical environments. Therefore, a hybrid approach that combines the efficiency of CNNs with temporal self-attention offers a promising solution.

This challenge is further compounded by noisy or irrelevant frequency components that hinder classification performance. Effective spectral feature selection is therefore critical, highlighting the need for a targeted approach that emphasizes relevant frequency bands to optimize both model accuracy and efficiency.

\subsection{Feature Importance-Based Feature Selection}

Feature selection is essential for improving machine learning performance, reducing dimensionality, and enhancing interpretability and efficiency. In tabular data, techniques such as decision trees, permutation feature importance \cite{widmann2024classifying}, Random Forests, and Recursive Feature Elimination (RFE) have effectively identified relevant features to boost classification accuracy \cite{chen2020selecting, aggarwal2023meticulous}. In medical imaging, meta-heuristic algorithms have been used to refine features extracted from models such as ResNet18 prior to classification with traditional classifiers like SVMs \cite{atban2023traditional}. Similarly, hybrid statistical methods have improved performance in high-dimensional domains such as Intrusion Detection Systems by reducing the feature space without compromising accuracy \cite{thakkar2023fusion}.

Moreover, feature selection within CNN-based audio classification, especially for respiratory sound spectrograms, remains underexplored. While a variety of techniques exist to assess feature importance in CNNs \cite{yang2023survey}, they are typically developed for visual explainability in natural image domains \cite{zhang2020measuring} and are often limited to qualitative analysis. Some recent works have attempted to leverage importance scores for selection \cite{sosa2024feature}, yet these do not address the unique spectral structure of respiratory sounds, where informative patterns frequently extend across continuous frequency bands rather than isolated bins.

To bridge this gap, we emphasize the need for frequency-band-level feature selection guided by deep learning and explainable AI.

\section{Proposed Method}

\label{sec:method}

\begin{figure*}
	\centering
	\includegraphics[height=0.4\textheight]{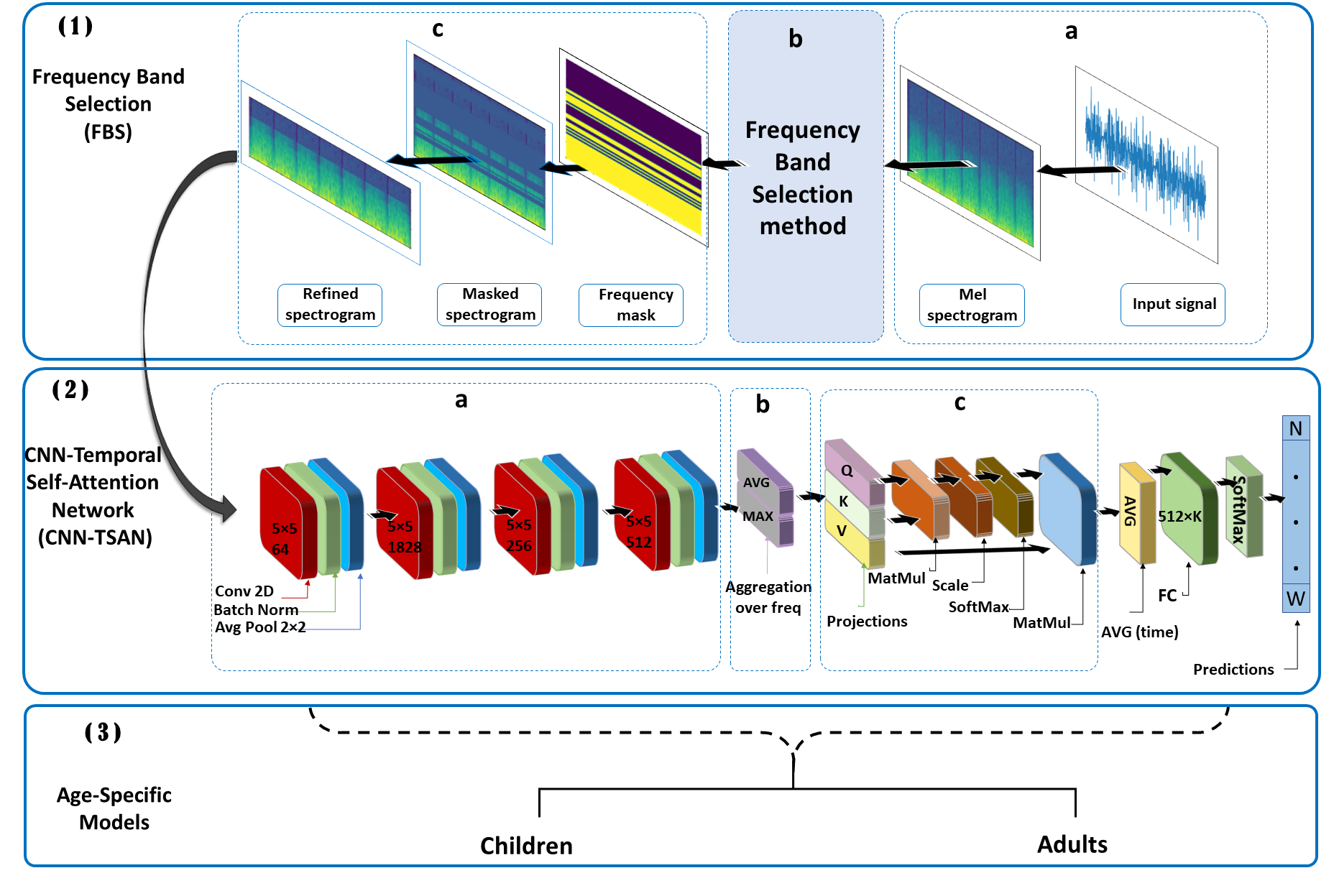}
	\caption{Overview of the proposed framework with three components: (1) Frequency Band Selection (FBS), transforming audio into Mel spectrograms, generating a frequency mask, and refining the input; (2) CNN-TSA, consisting of a CNN backbone, spectral aggregation, and temporal self-attention, followed by classification; and (3) Age-Specific Models, adapting classification for children and adults.}

	\label{FIG:1}
\end{figure*}

We introduce a respiratory sound classification framework consisting of three modules: FBS, CNN-TSA, and age-specific models, as illustrated in Figure~\ref{FIG:1}. 

In the FBS stage (Figure~\ref{FIG:1}-1), raw audio is first transformed into Mel spectrograms (Figure~\ref{FIG:1}-1(a)). Our importance-based FBS method then identifies and selects the most informative frequency bands (Figure~\ref{FIG:1}-1(b)). The resulting frequency mask is applied to the input spectrogram (Figure~\ref{FIG:1}-1(c)), enhancing feature relevance while reducing computational cost.

The masked spectrogram is then processed by the CNN-TSA module (Figure~\ref{FIG:1}-2), which consists of a CNN backbone (Figure~\ref{FIG:1}-2(a)), a frequency aggregation block (Figure~\ref{FIG:1}-2(b)), a temporal self-attention mechanism (Figure~\ref{FIG:1}-2(c)), and a final classifier.  The convolutional layers capture local spectral features, while the attention mechanism models long-range temporal dependencies. Moreover, to leverage the shared characteristics of the temporal cyclic patterns in lung sounds, aggregation is applied over the frequency dimension to preserve temporal patterns and reduce computational complexity.

To further improve accuracy, we introduce age-specific models (Figure~\ref{FIG:1}-3), which involve fine-tuning separate networks for adult and pediatric populations. This allows the model to account for physiological and acoustic differences across age groups.

The following subsections detail the CNN-TSA architecture and FBS methodology, providing deeper insight into the framework’s operation.

\subsection{CNN Temporal Self-Attention (CNN-TSA)}

The CNN-TSA network processes Mel spectrograms, which are computed from raw audio signals and serve as an effective representation of respiratory sounds~\cite{davis1980comparison}. The components of the network are detailed below.
\paragraph{CNN Backbone.}
The architecture begins with a CNN backbone comprising 2D convolutional layers, each followed by batch normalization and average pooling. These layers progressively downsample the input while extracting localized spectral features. The design is based on the CNN6 model from the PANNs (Pretrained Audio Neural Networks) framework, originally developed for large-scale audio tagging~\cite{kong2020panns}, offering a strong balance between accuracy and computational efficiency. To enhance feature extraction, the convolutional layers are initialized with weights pretrained on the AudioSet dataset.
For the ICBHI dataset, we adopt the full CNN6 configuration, which includes four convolutional layers and produces an output feature map with $d=512$ channels. For the SPRSound datasets, a three-layer variant is used—omitting the final convolutional layer, resulting in a feature map with $d=256$ channels. This modification is guided by empirical observations showing that a shallower architecture is more effective for SPRSound. In both settings, the output feature map preserves the temporal and frequency dimensions required for subsequent processing.

\paragraph{Aggregation Block}
To mitigate the high computational cost of self-attention, which scales quadratically with input size, we introduce an aggregation block (Figure~\ref{FIG:1}-2(b)) prior to the temporal self-attention module. This block reduces the feature map along the frequency axis by computing both average and max values across the frequency dimension of the CNN backbone's output. The average operation captures general spectral trends~\cite{kong2020panns}, while the max operation highlights prominent activations associated with anomalies such as wheezes and crackles, which often manifest as high-intensity peaks~\cite{piirila1995crackles}. The outputs of these two operations are summed, producing a compact yet informative representation that preserves both typical and salient features. This enables the self-attention mechanism to focus more effectively on temporal dynamics at a reduced computational cost.

\paragraph{Temporal Self-Attention} The temporal self-attention mechanism (Figure~\ref{FIG:1}-2(c)) is key to capturing long-range temporal dependencies. The attention mechanism operates on the aggregated feature matrix $\mathbf{X} \in \mathbb{R}^{T \times d}$, where $T$ is the number of time steps and $d$ represents the feature dimension, corresponding to the respective number of output channels from the aggregation block. The mechanism employs scaled dot-product attention~\cite{vaswani2017attention}, projecting the input $\mathbf{X}$ into query, key, and value spaces using learnable projection matrices:

\begin{equation}
\mathbf{Q} = \mathbf{X} \mathbf{W_Q}, \quad \mathbf{K} = \mathbf{X} \mathbf{W_K}, \quad \mathbf{V} = \mathbf{X} \mathbf{W_V},
\end{equation}
where $\mathbf{W_Q}, \mathbf{W_K} \in \mathbb{R}^{d \times d_k}$ and $\mathbf{W_V} \in \mathbb{R}^{d \times d}$ are learnable weight matrices, and $d_k = \frac{d}{8}$ denotes the reduced dimensionality of the key, chosen to optimize computational efficiency.

The attention output is computed as:

\begin{equation}
\text{Attention}(\mathbf{Q}, \mathbf{K}, \mathbf{V}) = \text{softmax}\left(\frac{\mathbf{QK}^T}{\sqrt{d_k}}\right) \mathbf{V}.
\end{equation}

Following attention, temporal aggregation is performed by averaging over the time dimension for each channel. This contrasts with flattening, which discards inter-channel structure and significantly increases parameters in downstream layers. Averaging preserves channel-wise characteristics while reducing each to a single scalar, enabling a compact and interpretable representation~\cite{kong2020panns, moummad2023pretraining}. The resulting vector is then passed to a fully connected linear layer for classification, significantly reducing model size while maintaining discriminative capacity~\cite{lin2014networknetwork}.

\subsection{Frequency Band Selection}

Noise and irrelevant features in respiratory sound data obscure essential information and impair model performance. To address this, we introduce a novel \emph{Importance-Based} frequency band selection strategy, applied on Mel spectrograms, where the horizontal axis represents time and the vertical axis corresponds to Mel frequency bands indexed as $f \in {1, \ldots, F}$, with $F$ denoting the total number of bands, each covering a distinct frequency range. Leveraging the fact that critical respiratory features are concentrated in specific frequency regions, the method iteratively identifies and prioritizes informative bands using model-driven feature attribution. This targeted refinement highlights relevant spectral components, resulting in improved classification accuracy, reduced input dimensionality, and enhanced computational efficiency and generalization.

Details of our proposed method are provided in Section~\ref{sec:importance_based}, while a standard backward selection technique is presented in Section~\ref{sec:backward_selection} for comparison and benchmarking purposes.

\subsubsection{Importance-Based Approach}
\label{sec:importance_based}

Each band $f$ receives an importance score that balances its mean contribution across classes with its class-wise stability, ensuring both relevance and consistency.

To ensure the robustness and generalizability of the attribution-based importance scores, K-fold cross-validation is employed on the training set.
For each fold $k$, we compute the pixel-wise attributions $\mathbf{A}_s^{c,k}$ for all spectrograms $s \in \text{trainset}_{c,k}$, which includes all samples labeled as class $c$ within the training set of fold $k$. The attribution matrix $\mathbf{A}_s^{c,k}$ is organized by frequency (rows) and time (columns), where $A_s^{c,k}(t,f)$ denotes the attribution of frequency band $f$ at time $t$.These attributions are derived using Grad-CAM \cite{selvaraju2017grad} within our CNN-TSA model.
a method shown to be robust for CNN-based architectures, though other attribution techniques may also be applied.
Grad-CAM computes the attribution map $\mathbf{A}_s^{c,k}$ by calculating the gradient of the class score ${\rm score}_c$ with respect to the feature maps ${\mathbf{M}_m}$ of the last CNN layer in the CNN-TSA model (where $\mathbf{M}_m$ denotes the $m$-th feature map), applying global average pooling to obtain importance weights, and then scaling and summing the feature maps. This process is mathematically expressed as:
\begin{equation}
\mathbf{A}_s^{c,k} = \sum_{m} \left(
        \frac{1}{H \times W} 
        \sum_{i=1}^{H}\sum_{j=1}^{W} 
        \frac{\partial {\rm score}_c}{\partial M_{m,ij}}
    \right)
    \mathbf{M}_m,
\end{equation}
where $H$ and $W$ are the spatial dimensions, $M_{m,ij}$ is the activation at location $(i,j)$, and $\frac{\partial {\rm score}_c}{\partial M_{m,ij}}$ measures the sensitivity of class score ${\rm score}_c$ to that activation. Both positive and negative contributions are preserved by omitting the ReLU step. The resulting $\mathbf{A}_s^{c,k}$ is rescaled using bilinear interpolation to match the dimensions of the input space. This yields the final attribution matrix $\mathbf{A}_s^{c,k}$ for each spectrogram $s$ of class $c$ in fold $k$.

The next step is to aggregate these attribution matrices to obtain a single attribution score per frequency band $f$ for class $c$ across the entire dataset. To compute the attribution of each frequency band f for class c, we first average $A_s^{c,k}(t,f)$ over all time steps, for each sample s, to obtain $A_s^{c,k}(f)$. We then compute the mean attribution by averaging $A_s^{c,k}(f)$ over all samples in trainset$_{c,k}$, resulting in $A^{c,k}(f)$.

After completing all folds, we average the class-specific mean attributions over folds to obtain $A^{c}(f)$ for each class $c$:
\begin{equation}
    A^{c}(f) = \frac{1}{K} \sum_{k=1}^{K} A^{c,k}(f),
\end{equation}

To quantify the overall contribution of each frequency band across all classes, we compute the average attribution. Specifically, for each frequency band $f$, we define:
\begin{equation}
\text{Mean}[f] = \frac{1}{|\mathcal{C}|} \sum_{c=1}^{|\mathcal{C}|}  A^{c}(f),
\label{eq:mean}
\end{equation}
where $\mathcal{C}$ is the set of classes and $|\mathcal{C}|$ its cardinality. This average reflects the overall reliance of the model on frequency band f. A higher $\text{Mean}[f]$ implies that band $f$ contributes significantly across many (or all) classes.

Beyond the mean contribution, it is essential to assess the consistency of each band's importance across classes. A band with high overall relevance may still be unreliable if its attribution varies significantly between classes. To promote uniform contribution and reduce class-specific bias, we quantify variability using the maximum inter-class attribution difference:

\begin{equation}
\text{MaxDiff}[f] = \max_{1 \leq c < c' \leq |\mathcal{C}|} |  A^{c}(f) -  A^{c'}(f) |.
\label{eq:maxdiff}
\end{equation}

A \emph{larger} value of $\text{MaxDiff}[f]$ indicates that the contribution of frequency band $f$ varies significantly across class pairs, reflecting inconsistency in its usage.

To compute a comprehensive importance score for each frequency band, we combine its mean contribution with a penalty for class-wise inconsistency. The resulting score, denoted as $\mathcal{I}_f$, is defined as:

\begin{equation}
\label{eq:Import}
\mathcal{I}_f = \text{Mean}[f] - \lambda \, \text{MaxDiff}[f],
\end{equation}
where $\lambda \in [0,1]$ is a hyperparameter that controls the penalty for class-wise variability; higher values prioritize consistency across classes, while lower values emphasize overall contribution.
With the importance metric $\mathcal{I}_f$ defined in Equation~\eqref{eq:Import}, we are now able to rank frequency bands based on their overall contribution and consistency across classes. However, directly removing all low-importance bands in a single step is not effective, as the model may be biased by the presence of noisy or irrelevant frequencies during training. This can obscure the true value of some bands and degrade overall performance.

To address this, we introduce a hyperparameter $r$ that determines the number of frequency bands to eliminate in each iteration. This iterative elimination process, detailed in Algorithm~\ref{algo}, allows the model to gradually adapt and focus on more informative spectral regions. In each iteration, 5-fold cross-validation is used to compute updated importance scores, the $r$ least important bands are removed, and the model is retrained with the refined frequency subset. This progressive refinement enables the discovery of interactions among the remaining bands and improves model robustness. The process continues until a predefined stopping criterion is met, in this case, a drop in performance. Once the optimal frequency subset is identified, the final model is retrained on the entire training set and evaluated on the official test sets of each dataset (see Section~\ref{sec:datasets} for dataset details and splits).

\begin{algorithm}[t]
\caption{Iterative Importance-Based Frequency Band Selection}
\label{algo}
\KwIn{Dataset $D = \{\text{Training}, \text{Testing}\}$; class set $C$; model $\mathcal{N}$; hyperparameter $\lambda$; bands removed per iteration $r = 4$}
\KwOut{Optimized frequency mask $\mathbf{m}$}

Train $\mathcal{N}$ on the training set using K-fold cross-validation (CV)\;

\While{no degradation in CV performance}{
    \ForEach{class $c \in C$}{
        Compute mean attribution $ A^{c}(f)$ across all training spectrograms of class $c$\;
    }
    Compute importance scores $\mathcal{I}_f$ (eq~\ref{eq:Import})\;

    Eliminate the $r$ bands with the lowest $\mathcal{I}_f$\;
    Update the binary mask $\mathbf{m}$;

    Retrain $\mathcal{N}$ using $\mathbf{m}$ and evaluate with CV\;
}

Train and evaluate $\mathcal{N}$ on $D$ using the final mask $\mathbf{m}$\;
\end{algorithm}

\subsubsection{Backward Selection Approach}
\label{sec:backward_selection}
The backward selection approach aims to identify an optimal subset of frequency bands by iteratively removing the least informative ones. Unlike traditional one-band-at-a-time strategies, we remove four adjacent bands per iteration, leveraging harmonic similarities among neighboring frequencies~\cite{marques2018normal} to accelerate convergence.

At each iteration, K-fold cross-validation is conducted on the training set. Each candidate subset (four adjacent bands) is tentatively excluded, and the configuration resulting in the least performance degradation or the greatest improvement is selected for removal. This process continues until a significant drop in validation accuracy indicates that further pruning would be detrimental. The final model is trained on the full training set using the selected frequency subset and evaluated on the test set. 

Standard backward selection has a computational complexity of $\mathcal{O}(F^2)$. The grouped-band variant reduces the number of candidates per iteration by evaluating groups of four adjacent bands, lowering the cost to $\mathcal{O}((F/4)^2)$, but still retains a quadratic relationship. In contrast, our importance-based method (Section~\ref{sec:importance_based}) achieves a more efficient linear complexity of $\mathcal{O}(F)$ by computing attribution scores once per iteration and directly removing the four least important bands, avoiding exhaustive evaluation.

\subsection{Adaptation of Frequency Band Selection to Transformer-Based Models}

Although our Frequency Band Selection (FBS) method was originally developed for the CNN-TSA architecture, its model-agnostic design allows it to be extended to other neural network paradigms. To demonstrate its generalizability, we applied FBS to a transformer-based approach from the Patch-Mix Contrastive Learning framework for respiratory sound classification~\cite{bae2023patch}.

The overall FBS pipeline was preserved as described in Section~\ref{sec:importance_based}. However, a key adaptation involves the attribution method used to guide the importance-based selection. Since Grad-CAM relies on spatial activation maps, we employed Integrated Gradients (IG)~\cite{sundararajan2017axiomatic}. IG offers a robust alternative by quantifying feature importance through the integration of the gradients of the model’s prediction with respect to the input features along a path from a baseline to the actual input. This makes it particularly appropriate for transformer models, where attention operates over input patches rather than localized convolutional maps.

This adaptation demonstrates the flexibility of FBS and its potential to enhance both performance and efficiency across a variety of model architectures.

\section{Experimental settings }

\label{sec:experimental settings}

\subsection{Datasets and Proposed Tasks}
\label{sec:datasets}

To evaluate the effectiveness of our proposed method for respiratory sound classification, we utilized two well-established datasets: the ICBHI 2017~\cite{rocha2019open} and SPRSound~\cite{zhang2022sprsound}.


\subsubsection{ICBHI 2017 Dataset}

The ICBHI 2017 dataset~\cite{rocha2019open} comprises 920 respiratory sound recordings collected from 126 participants aged 0.25 to 93 years. Recordings were acquired using four types of stethoscopes, 8AKGC417L, Meditron, Litt3200, and LittC2SE, with sampling rates ranging from 4,000 Hz to 44,100 Hz and durations between 10 and 90 seconds. Each recording contains multiple breathing cycles, segmented and annotated to indicate the presence of adventitious sounds. The dataset includes 3,642 normal cycles, 1,864 with \textit{crackles}, 886 with \textit{ wheezes}, and 506 with both \textit{crackles and wheezes}. Cycle durations range from 0.2 to 16 seconds, with an average of 2.7 seconds.

The official data split assigns 60\% of recordings to training and 40\% to testing, with no subject overlap between sets.

\subsubsection{SPRSound Dataset}

The SPRSound datasets, introduced in 2022 and expanded in 2023, comprise a comprehensive collection of pediatric respiratory sound recordings for the classification of adventitious sounds. The SPRSound 2022 dataset~\cite{zhang2022sprsound} includes 2,683 recordings with 9,089 annotated sound events from 292 children aged 0.2 to 16.2 years, captured at 8 kHz using a digital stethoscope. Events are labeled into seven categories: \textit{Normal}, \textit{Rhonchi}, \textit{Wheeze}, \textit{Stridor}, \textit{Coarse Crackle}, \textit{Fine Crackle}, and \textit{Wheeze and Crackle}. For the BioCAS Grand Challenge 2022~\cite{zhang2022grand}, the dataset is split into a training set of 1,949 recordings (6,656 events) from 251 participants and a test set of 734 recordings (2,433 events), supporting both intra- and inter-patient evaluation. The SPRSound 2023 dataset~\cite{zhang2023grand} extends this benchmark with 871 test-only recordings and 3,124 events from 95 pediatric subjects, maintaining the same event taxonomy. Together, these datasets enable rigorous assessment of pediatric respiratory sound classification methods.

Two classification tasks are defined :\\
 \textbf{Task 1: Multi-class Classification}: Classify each respiratory cycle into one of the annotated sound classes (e.g., \textit{normal}, \textit{crackles}, \textit{wheezes}, etc.).\\
\textbf{Task 2: Binary Classification}: Distinguish between \textit{normal} and \textit{adventitious} lung sounds, where all abnormal classes are grouped as adventitious.
  
\subsection{Evaluation Metrics}
\label{sec:metrics}

In alignment with the evaluation metrics of the ICBHI 2017 challenge \cite{rocha2019open} and SPRSound challenges \cite{zhang2022sprsound}, we employ the following metrics to evaluate our framework: sensitivity ($S_e$), specificity ($S_p$), average score ($AS$), harmonic score ($HS$), and total score ($TS$).

\begin{equation}
\begin{aligned}
S_e &= \frac{\text{\# correctly identified adventitious sounds}}{\text{\# total adventitious sounds}} \\
S_p &= \frac{\text{\# correctly identified normal sounds}}{\text{\# total normal sounds}} \\
{\rm AS} &= \frac{S_e + S_p}{2}, \quad
{\rm HS} = \frac{2 \cdot S_e \cdot S_p}{S_e + S_p} \\
{\rm TS} &= \frac{{\rm AS} + {\rm HS}}{2}
\end{aligned}
\end{equation}

For the age-specific models, the reported performance metric is the average of both models 

\subsection{Preprocessing details}
The audio recordings vary in sampling rates and durations. To standardize them, all recordings are re-sampled to 16~kHz mono. Following prior work \cite{moummad2023pretraining,song2021contrastive,nguyen2022lung}, each respiratory cycle is limited to a maximum of 8 seconds. Shorter clips are extended using circular padding or repetition with truncation and fade-out for smooth transitions; longer ones are simply truncated. Audio signals are transformed into Mel-spectrograms using 64 Mel filter banks \cite{nakamura2016detection,xu2021arsc,moummad2023pretraining}, with a window size of 1024, hop size of 512, and frequency range of 50–2000~Hz to cover the typical bands of wheezes and crackles \cite{jakovljevic2018hidden}. SpecAugment \cite{park2019specaugment} is applied to mitigate overfitting
. For the adaptation of our Frequency Band Selection (FBS) method within the Patch-Mix Contrastive Learning framework, we adopt the preprocessing pipeline from \cite{bae2023patch}, which also uses Mel-spectrogram representations.

\subsection{Training Configuration and Implementation Details}
To evaluate our system on both binary and multiclass classification tasks, we used the official train-test splits provided with each dataset. For frequency band selection, a 5-fold cross-validation was conducted within the training set, ensuring patient-wise separation between training and validation partitions.

To address class imbalance, we employed the weighted categorical cross-entropy (WCCE) loss function \cite{ho2019real}, which assigns class weights inversely proportional to their frequencies, promoting balanced learning across all classes:

\begin{equation}
 L_{WCCE} = -\frac{1}{N} \sum_{c=1}^{|\mathcal{C}|} \sum_{n=1}^{N} w_c y_n^c\log(\mathcal{N}_{\boldsymbol{\theta}}(\mathbf{x}_n, c)),   
\end{equation}
where $N$ is the total number of samples, $w_c$ is the class weight defined as the inverse of the number of samples in class $c$, $y_n^c$ is the ground truth label, $\mathbf{x}_n$ is the input, and $\mathcal{N}_{\boldsymbol{\theta}}$ is the neural network which is parameterized with parameter vector $\boldsymbol{\theta}$.

Training was performed for 200 epochs on both the ICBHI and SPRSound datasets using the Adam optimizer (momentum = 0.9), a batch size of 128 (reduced to 64 for age-specific models), an initial learning rate of $10^{-3}$, and a weight decay of $10^{-4}$. A cosine annealing schedule was used for learning rate decay. For experiments involving the proposed Frequency Band Selection (FBS) within the Patch-Mix Contrastive Learning framework \cite{bae2023patch}, we adopted the same hyperparameter settings reported in the original study.

\section{Experimental results  }

\label{sec:experimental results}

\subsection{Ablation study}
To assess the effectiveness of the proposed methods, we conducted a series of experiments to evaluate the contributions of each component and methodological decision.  

\subsubsection{Impact of Temporal Self-Attention Block Placement}

Table \ref{tab:attn-placement} presents the effect of Temporal Self-Attention (TSA) placement within the CNN backbone on the trade-off between computational cost (measured in GFLOPs) and model performance (measured by Average Score, AS) for Task-1 on the ICBHI 2017 dataset. The baseline model attains an AS of 55.55\% at 2.47 GFLOPs. Introducing TSA at the input level results in a performance drop and a slight increase in cost, suggesting that early attention on raw features may amplify irrelevant signal variations. Placing TSA after the first or second convolutional blocks yields AS scores of 54.83\% and 55.30\%, respectively, with increased costs (3.17 and 2.81 GFLOPs), indicating limited benefit and potential disruption of feature hierarchies. Improved performance is observed when TSA is inserted after the third convolutional block, leveraging more abstract temporal representations. TSA placement after the final convolutional block achieves 57.20\% at 2.55 GFLOPs, balancing accuracy and efficiency. The highest performance (58.08\%) with a reduced computational cost (2.48 GFLOPs) is achieved when TSA is applied post-aggregation, enabling effective attention over temporally aligned, frequency-aggregated representations. These findings underscore the critical role of TSA placement, with post-aggregation integration yielding optimal performance and efficiency.

\begin{table}[!h]
\centering
\caption{Impact of TSA placement in the CNN backbone on GFLOPs and Average Score (AS) for Task-1 on the ICBHI 2017}
\label{tab:attn-placement}
\begin{tabular}{lcc}
\hline
\textbf{Method} & \textbf{$AS$  (\%)} & \textbf{GFLOPs} \\
\hline
Baseline (CNN) & 55.55 & 2.47 \\
TSA on Input & 52.35 & 2.49 \\
TSA after 1st Conv Block & 54.83 & 3.17 \\
TSA after 2nd Conv Block & 55.30 & 2.81 \\
TSA after 3rd Conv Block & 56.86 & 2.64 \\
TSA after Last Conv Block & 57.20 & 2.55 \\
\textbf{TSA after Aggregation} & \textbf{58.08} & \textbf{2.48} \\
\hline
\end{tabular}
\vspace{-2mm}
\end{table}

\subsubsection{Performance Across Iterative Frequency Band Elimination}

\begin{figure}[!h]
    \centering
    \includegraphics[width=1\linewidth]{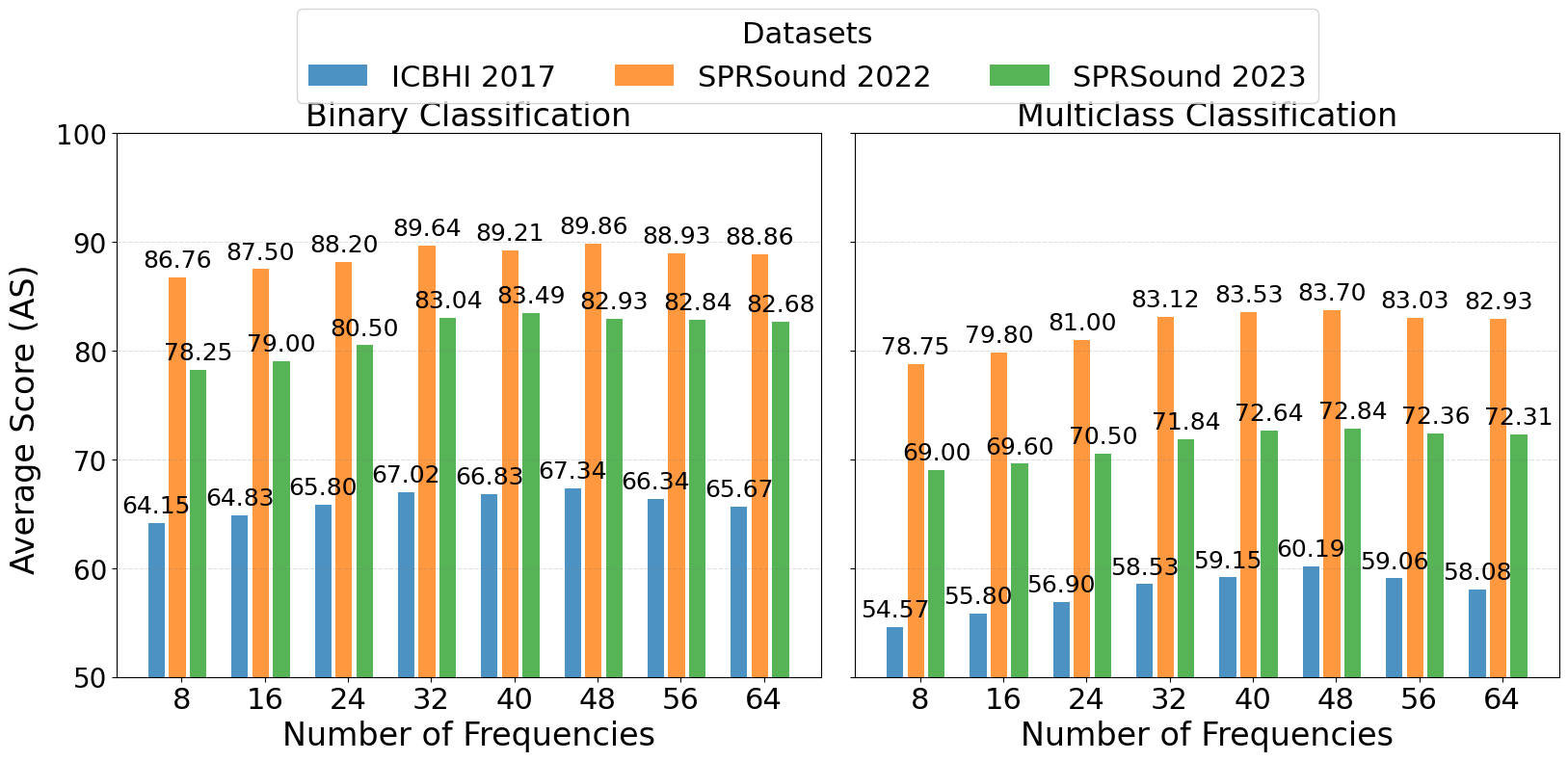 }
    \caption{Average Scores (AS) for Binary (left) and Multiclass (right) Classification across ICBHI 2017, SPRSound 2022, and SPRSound 2023 datasets using different numbers of retained frequency bands.}
    \label{fig:P}
\end{figure}

We evaluated the effect of iterative frequency band elimination using our importance-based selection method (FBS (IS)) with the CNN-TSA model across Binary and Multiclass Classification tasks on the ICBHI 2017, SPRSound 2022, and SPRSound 2023 datasets.As shown in Figure~\ref{fig:P}, performance was recorded at intervals of 8 bands, while the least important bands were removed in steps of 4 from the full set of 64, down to 8 bands.

For Binary Classification, performance improved as less informative bands were removed, peaking around 48 bands. On ICBHI 2017 and SPRSound 2022, AS increased from 65.67\% and 88.86\% to 67.34\% and 89.86\%, respectively. SPRSound 2023 reached a peak AS of 83.49\% at 40 bands. Additionally, the $AS$ score with 32 bands exceeded that of the full spectrum (64 bands) across all datasets, confirming the effectiveness of our method.

In Multiclass Classification, a similar pattern was observed, with optimal performance at 48 bands. AS increased from 58.08\% to 60.19\% on ICBHI 2017, and SPRSound 2022 and 2023 peaked at 83.70\% and 72.84\%, respectively. Performance remained comparable with the full spectrum, even with 32–40 bands, demonstrating that substantial frequency reduction can be achieved with minimal loss of accuracy. These findings underscore the value of informed frequency band selection for improving efficiency while preserving or enhancing classification performance.

\subsubsection{Impact of the Hyperparameter $\lambda$ on Model Performance}

\begin{figure}[!h]
    \centering
    \includegraphics[width=1\linewidth]{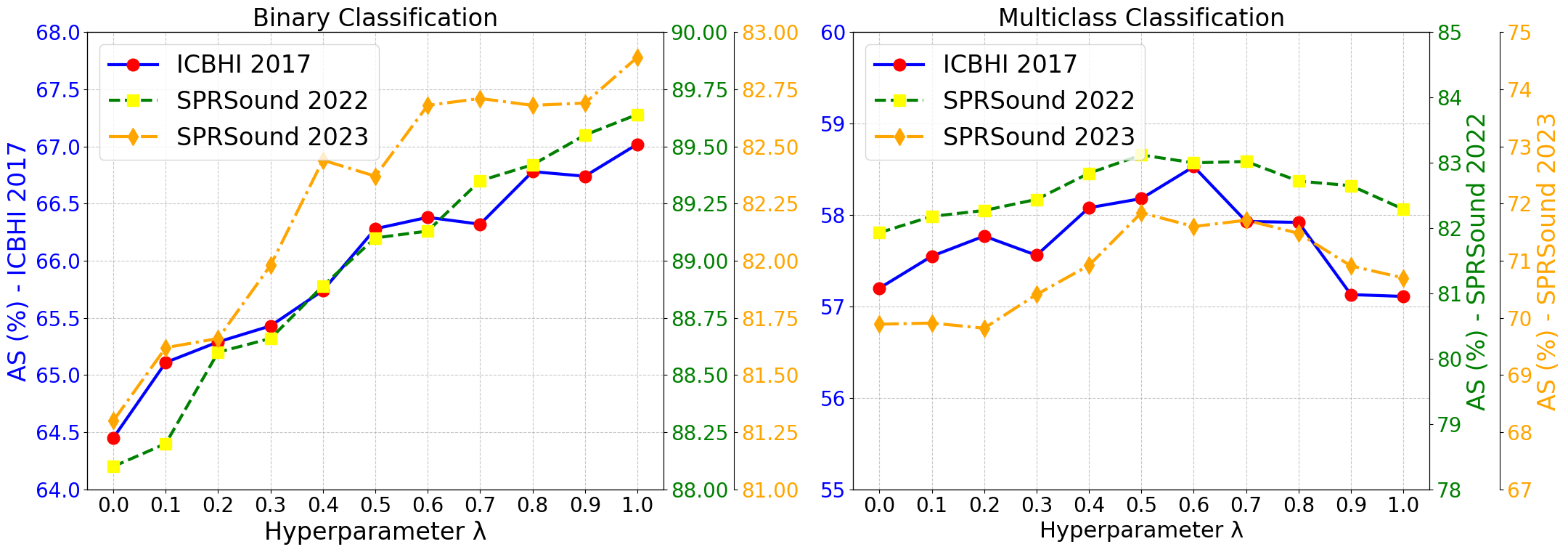}
    \caption{Impact of the hyperparameter $\lambda$ on model performance.
The plot shows the Average Score ($AS$)  across
different values of $\lambda$ }
    \label{fig:lambda}
\end{figure}

Figure~\ref{fig:lambda} illustrates how the weighting hyperparameter $\lambda$ affects classification performance across the ICBHI 2017, SPRSound 2022, and SPRSound 2023 datasets using our importance-based frequency band selection (FBS (IS)) with 50\% of bands retained. The importance score is defined as $\mathcal{I}_f = \text{Mean}[f] - \lambda \cdot \text{MaxDiff}[f]$, where $\lambda \in [0, 1]$ controls the trade-off between overall contribution and class-specific variability.

In Binary Classification, performance consistently improves with higher $\lambda$, indicating that emphasizing globally informative bands enhances generalization. AS increases on all datasets, with ICBHI 2017 reaching 67.02\%, SPRSound 2022 reaching 89.64\%, and SPRSound 2023 reaching 82.89\% at $\lambda = 1.0$.

For Multiclass Classification, peak performance occurs at intermediate $\lambda$ values (0.5–0.6), where both shared and class-distinctive frequency bands are leveraged. ICBHI 2017 peaks at 58.53\%, SPRSound 2022 at 83.12\%, and SPRSound 2023 at 71.84\%.

These results show that higher $\lambda$ values favor generalization in binary tasks, while moderate values enable better discrimination in multiclass settings by balancing global and class-specific features.

\subsubsection{Analysis of Frequency Masks: Backward vs. Importance-Based Selection on ICBHI 2017}

\begin{figure}[!h]
    \centering
    \includegraphics[width=0.9\linewidth]{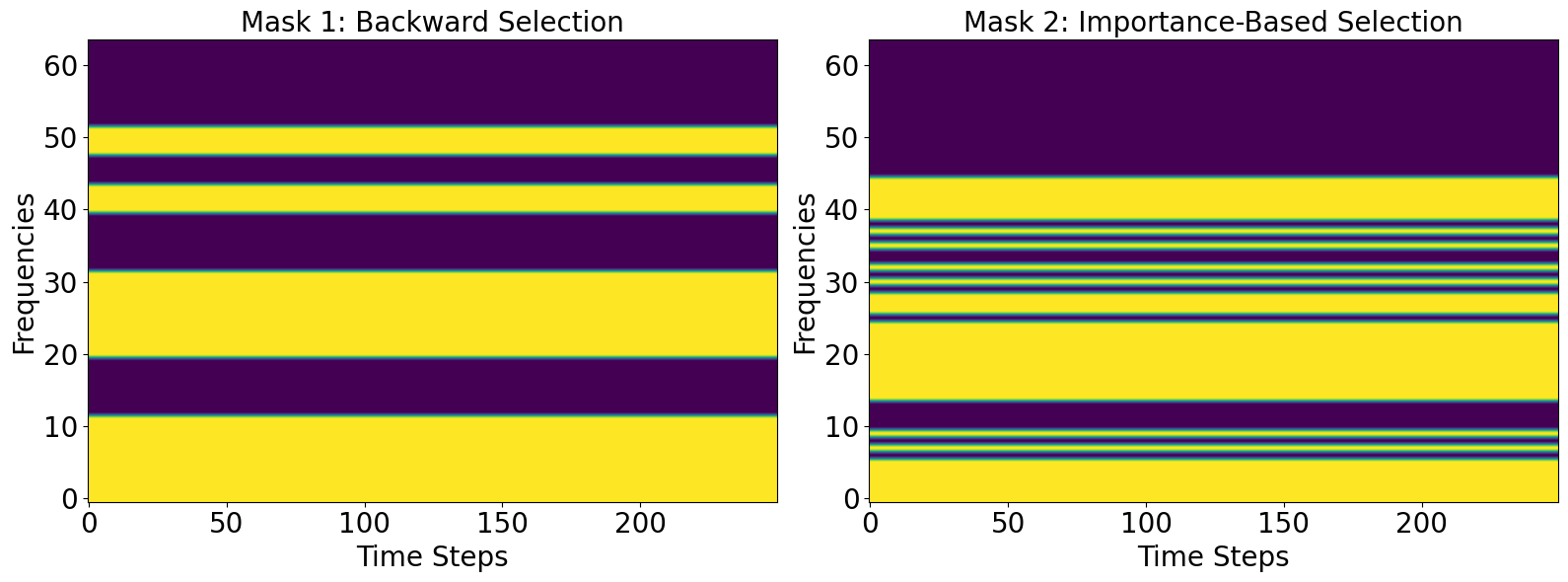}
   \caption{Binary masks from backward selection (left) and importance-based selection (right).}
    \label{fig:BS vs IS}
\end{figure}

Figure~\ref{fig:BS vs IS} compares the binary frequency masks produced by the backward selection method (left) and the importance-based selection method (right), both removing 4 bands per iteration for consistency. The backward selection approach, based solely on performance degradation, leads to broader, less targeted band elimination because the 4 eliminated bands need to be adjacent to reduce the number of combinations. In contrast, the proposed importance-based method yields a more refined and structured mask by considering both the overall contribution and class-wise consistency of each band. This results in a more compact, interpretable, and discriminative representation, improving both model efficiency and classification performance.

\subsubsection{Ablation Study on ICBHI 2017 Dataset}

Table~\ref{tab: ablation study on ICBHI} presents the performance and efficiency of various system configurations on the ICBHI 2017 dataset. The baseline CNN model achieved an AS of 55.55\% with 2.47 GFLOPs. Adding temporal self-attention (CNN-TSA) improved performance to 58.08\% with a negligible increase in computational cost.
Incorporating frequency band selection (FBS) led to substantial efficiency gains. Both the backward (FBS(BS)) and importance-based (FBS(IS)) methods reduced GFLOPs by up to 50\%, while achieving better performance, compared to CNN-TSA alone.
In child-specific models (C), the baseline performed better with an AS of 64.21\%, reflecting lower inter-patient variability. These models further benefited from both TSA and FBS, with the FBS(BS) configuration achieving the highest AS of 67.32\%. When combining adult (A) and child (C)models, CNN-TSA with FBS(BS) achieved the best AS of 62.68\%, with FBS(IS) performing comparably.
These results highlight the effectiveness of combining temporal self-attention with informed frequency band selection, particularly the importance-based variant, in improving classification performance and computational efficiency.
\begin{table}[!htbp]
\centering
\caption{Results analysis on ICBHI concerning GFLOPs, parameters, and performance metrics}
\label{tab: ablation study on ICBHI}
\begin{tabular}{>{\raggedright\arraybackslash}p{3.5cm} |c| c| c| c}
\hline
\textbf{Method} & \text{GFLOPs} & \textbf{$S_p$(\%)} & \textbf{$S_e$(\%)} & \textbf{$AS$(\%)} \\ \hline
Baseline (CNN) & 2.47 &  73.21 & 37.89 & 55.55 \\ \hline
CNN-TSA & 2.48 & 78.78 &  37.38 & 58.08 \\\hline
CNN-TSA+FBS(BS)$_{(50\%freq)}$ & 1.24 &  79.00 & 37.48 & 58.24  \\ \hline
CNN-TSA+FBS(IS)$_{(50\%freq)}$ & 1.24 &  78.53 & 38.49 & 58.51 \\ \hline
Baseline (A) & 2.47 &  75.97 & 36.37 &  56.17\\ \hline
CNN-TSA (A) & 2.48 & 77.63 & 36.89 &57.27  \\ \hline
CNN-TSA+FBS(BS)$_{(50\%freq)}$ (A) & 1.24 &  75.50 & 40.56 & 58.03\\ \hline
CNN-TSA+FBS(IS)$_{(50\%freq)}$ (A) & 1.24 &  70.06 & 44.85 & 57.46\\ \hline
Baseline (C) &  2.47& 87.76 & 40.66 & 64.21 \\ \hline
CNN-TSA (C) & 2.48 &  90.11 & 40.00 &  65.05\\ \hline
CNN-TSA+FBS(BS)$_{(50\%freq)}$ (C) & 1.24 &  93.97 & 40.67 & 67.32\\ \hline
CNN-TSA+FBS(IS)$_{(50\%freq)}$  (C) & 1.24 &  88.81 & 45.00 & 66.91 \\ \hline
Age-specific CNN-TSA &  2.48&   83.87& 38.44 & 61.16\\ \hline
Age-specific CNN-TSA+FBS(BS)$_{(50\%freq)}$ & 1.24 &   84.74 & 40.62 & 62.68\\\hline
Age-specific CNN-TSA+FBS(IS)$_{(50\%freq)}$  & 1.24 &   79.44 & 44.93 & 62.19\\ \hline
\end{tabular}
\end{table}

 \subsection{Comparative Analysis with State-of-the-Art Methods}
We evaluated the performance of our proposed method against various state-of-the-art approaches on the ICBHI 2017, SPRSound 2022, and SPRSound 2023 databases. For the ICBHI database, we employed the Age-specific models to address the significant variability in patient age. The results represent the mean and standard deviation over five runs.

Table~\ref{tab:performance_comparison on icbhi 2017} presents the performance comparison of our proposed methods, CNN-TSA and FBS variants, with state-of-the-art methods for 4-class (Task 1) and 2-class (Task 2) evaluations on ICBHI2017.

\begin{table*}[h]
\centering

\caption{Performance comparison between the proposed system and previous studies from the existing literature on the ICBHI 2017 database. The \textbf{best} results are highlighted in bold, and the \underline{second best} results are underlined.}
\label{tab:performance_comparison on icbhi 2017}
\begin{tabular}{lcccccc}
\hline
\textbf{Task} &\textbf{method} & \textbf{architecture} & \textbf{\#parameters} & \textbf{$S_p$ (\%)} & \textbf{$S_e$ (\%)} & $\textbf{AS} (\%)_{\pm{std}} $\\
\hline
\parbox[t]{2mm}{\multirow{28}{*}{\rotatebox[origin=c]{90}{Task1: 4-class eval.}}} 

&Yang et al. \cite{yang2020adventitious} ( SE+SA) & ResNet18 & 12M & 81.25 & 17.84 & 49.55 \\

&Li et al. \cite{li2021lungattn} (LungAttn) & ResNet-Att & 0.7 M& 71.44 & 36.36 & 53.90 \\
&Xu et al. \cite{xu2021arsc} (ARSC-Net) & bi-ResNet-Att & - & 67.13 & \textbf{46.38} & 56.76 \\
&Gairola et al. \cite{gairola2021respirenet} (RespireNet) & ResNet34 &  21M & 72.30 &40.10 & 56.20 \\
&Ren et al. \cite{ren2022prototype} & CNN8-PT & - & 72.96 & 27.78 & 50.37 \\
&Chang et al. \cite{chang22h_interspeech} & CNN8+dilated & - & 69.92 & 35.85& 52.89 \\
&Chang et al. \cite{chang22h_interspeech} & ResNet-dilated & - & 50.22 & 51.83 & 51.02 \\
&Wang et al. \cite{wang2022domain} (Splice) & ResNeSt & 25M & 70.40 & 40.20 & 55.30 \\
&Pham et al. \cite{pham2022ensemble} (Late-Fusion) & Inc-03 + VGG14 & - & \textbf{85.60} & 30.00 & 57.30 \\

&Nguyen et al. \cite{nguyen2022lung} (CoTuning) & ResNet50 & 23M & 79.34 & 37.24 & 58.29 \\
&Moummad et al. \cite{moummad2023pretraining} (SCL) & CNN6 & 4.3M & 75.95 & 39.15 & 57.55 \\

& Bae et al.\cite{bae2023patch}(Patch-Mix CL) & AST & 86M & 81.66 & 43.07 & 62.37 \\

&Niizumi et al.(M2D-X/0.7)\cite{niizumi2024masked}&AST&86M&81.51 & 45.08 &63.29\\
&Kim et al. (SG-SCL) \cite{kim2024stethoscope} & AST& 86M& 79.87 &  43.55 & 61.71  \\ 

&Kim et al.(BTS)\cite{kim2024bts}&CLAP & 190.8M & 81.40 & \underline{45.67}  & \underline{63.54}  \\ \cline{2-7}
&\textbf{Baseline (Ours)}& CNN& 4.3M & 73.21 &37.89  & $55.55_{\pm {0.8}} $\\
&\textbf{CNN-TSA  (Ours)}& CNN-TSA & 4.6M & 78.78 & 37.38 & $58.08_{\pm {0.77}}$
 \\
&\textbf{CNN-TSA +FBS(BS)$_{50\%freq}$(Ours)}& CNN-TSA  & 4.6M & 79.00 & 37.48 & $58.24_{\pm {0.9}}$\\
&\textbf{CNN-TSA +FBS(BS)$_{75\%freq}$(Ours)}& CNN-TSA  & 4.6M & 78.14 & 38 & $58.07_{\pm {0.88}}$ \\
&\textbf{CNN-TSA +FBS(IS)$_{50\%freq}$(Ours)}& CNN-TSA  & 4.6M &  78.53 & 38.49 & $58.51_{\pm {0.49}}$\\
&\textbf{CNN-TSA +FBS(IS)$_{75\%freq}$(Ours)}& CNN-TSA  & 4.6M &  80.37 & 40 & $60.19_{\pm {0.65}}$\\
&\textbf{Age-specific CNN-TSA +FBS(BS)$_{50\%freq}$(Ours)}& CNN-TSA  &4.6M & \underline{84.74} & 40.62 & $62.68_{\pm {1.07}}$\\
&\textbf{Age-specific CNN-TSA +FBS(IS)$_{50\%freq}$(Ours)}& CNN-TSA  &4.6M  & 79.44 & 44.93 & $62.19_{\pm {1.1}}$ \\
&Patch-Mix CL \cite{bae2023patch} + \textbf{FBS(IS)$_{50\%freq}$ (Ours)} & AST & 86M & 82.44 &42.74 & $62.59_{\pm {0.5}}$ \\
&Patch-Mix CL \cite{bae2023patch} + \textbf{FBS(IS)$_{75\%freq}$ (Ours)} & AST & 86M & 84.35 & 43.67 & $\textbf{64.01}_{\pm {0.66}}$ \\

\hline
\parbox[t]{2mm}{\multirow{10}{*}{\rotatebox[origin=c]{90}{Task2: 2-class eval.}}} 
&Pham et al. \cite{pham2021cnn} (CNN-MoE) & C-DNN & - & 72.40 & 37.50 & 54.10 \\

&Nguyen et al. \cite{nguyen2022lung} (CoTuning) & ResNet50 & 23M & 79.34 & 50.14 & 64.74 \\
& Bae et al.\cite{bae2023patch}(Fine-tuning) & AST & 86 M & 77.14 & 56.40 & 66.77 \\
&Bae et al.\cite{bae2023patch}(Patch-Mix CL) & AST & 86 M & \textbf{81.66} & 55.77 & 68.71\\

&Kim et al.\cite{kim2024stethoscope}(DAT) & AST & 86 M & \underline{79.87} & 57.97 & \underline{68.93}\\

\cline{2-7}
&\textbf{Baseline  (Ours)}& CNN& 4.3M & 62.38 & \textbf{66.35} & $64.37_{\pm {0.91}}$ \\
&\textbf{CNN-TSA (Ours)}& CNN-TSA & 4.6M & 67.87  & 62.57 & $65.22_{\pm {0.8}}$ \\
&\textbf{CNN-TSA +FBS(BS)$_{50\%freq}$(Ours)}& CNN-TSA  & 4.6M & 70.99 & 61.77 & $66.38_{\pm {0.88}}$\\
&\textbf{CNN-TSA +FBS(BS)$_{75\%freq}$(Ours)}& CNN-TSA  & 4.6M & 71.94 & 61.43   & $66.69_{\pm {1}}$\\
&\textbf{CNN-TSA +FBS(IS)$_{50\%freq}$(Ours)}& CNN-TSA  & 4.6M & 67.76 &  \underline{66.27}& $67.02_{\pm {0.75}}$\\
&\textbf{CNN-TSA +FBS(IS)$_{75\%freq}$(Ours)}& CNN-TSA  & 4.6M & 72.22 &  62.45& $67.34_{\pm 0.69}$\\
&\textbf{Age-specific CNN-TSA +FBS-BS$_{50\%freq}$(Ours)}& CNN-TSA  &4.6M  & 75.23 & 61.67 & $68.45_{\pm {1.1}}$ \\
&\textbf{Age-specific CNN-TSA +FBS(IS)$_{50\%freq}$(Ours)}& CNN-TSA  &4.6M  & 76.63 & 60.43 & $68.53_{\pm {1.03}}$ \\

&Patch-Mix CL \cite{bae2023patch} + \textbf{FBS(IS)$_{50\%freq}$(Ours)} & AST & 86M & 78.53
 & 58.58 &$68.56_{\pm {0.77}}$
 \\
&Patch-Mix CL \cite{bae2023patch} + \textbf{FBS(IS)$_{75\%freq}$(Ours)} & AST & 86M & 75.17
 & 65.0
 & $\textbf{70.08}_{\pm {0.57}}$
 \\
\hline
\end{tabular}
\end{table*}
For Task 1, our CNN-TSA + FBS(IS) model, with only 4.6M parameters, outperforms all CNN-based methods, including RespireNet (21M) and ResNet50 (23M), achieving an AS of 58.51\% using only 50\% of the frequency bands and 60.19\% when retaining 75\%. Additionally, the age-specific CNN-TSA + FBS(BS) model reaches an AS of 62.68\%, comparable to AST-based models such as Patch-Mix CL (86M) and Audio-CLAP (190M), despite its significantly smaller architecture. This demonstrates the efficiency of our approach in delivering competitive performance with a lightweight model.

We also integrated our FBS (IS) with the Patch-Mix CL framework, which includes an AST backbone that achieves state-of-the-art performance of 62.37\% on the ICBHI dataset. This integration resulted in an improved AS performance of 64.01\%, thereby setting a new state of the art and demonstrating that our method is also generalizable to other backbone architectures.

In the 2-class evaluation, the CNN-TSA + FBS(IS) model achieved an AS of 67.34\%, surpassing all CNN-based baselines and demonstrating competitive results with AST-based models like Patch-Mix CL. The age-specific CNN-TSA + FBS(IS)variant achieved an AS of 68.53\%, highlighting its robustness and generalization capability in binary classification. Moreover, applying FBS(IS) within the Patch-Mix CL framework resulted in the highest score of 70.08\%, setting again a new state-of-the-art performance for this task.

Table~\ref{tab:performance_comparison on biocas 2022/2023} compares the performance of our proposed models with existing methods on the SPRSound 2022 and SPRSound 2023 databases for Task 1 and Task 2.
For Task 1, the proposed CNN-TSA +FBS(IS) achieves the highest Total Score (TS) of 83.35\%, outperforming prior methods, including DenseNet169 \cite{ma2022effective} and ResNet-18 \cite{li2022improving}, which utilize significantly larger architectures (12M–15M parameters). This highlights the efficiency of our model, which achieves superior performance with only 1.11M parameters. Additionally, the frequency band selection mechanism reduces computational costs. For Task 2, the CNN-TSA +FBS(IS) model achieves the best TS (89.85\%), surpassing previous methods such as ResNet-18 with Focal Loss \cite{li2022improving} and CNN with MFCC \cite{babu2022multiclass}. These results demonstrate the robustness of the proposed approach in capturing critical features for binary classification tasks.

On the SPRSound 2023 database, our models also outperform previous methods in both Task 1 and Task 2. For Task 1, the CNN-TSA +FBS(IS) achieves the highest TS of 71.20\%, surpassing architectures like ResNet (SCL) and Inception-residual \cite{ngo2023deep}, while maintaining a significantly smaller parameter size. Similarly, for Task 2, the CNN-TSA +FBS(IS) model achieves the highest TS of 83.20\%, outperforming all competing methods.

\begin{table*}[h]
\centering
\caption{Performance comparison between the proposed system and previous studies from the existing literature on the SPRSound  2022 and SPRSound  2023 databases. The \textbf{best} results are highlighted in bold. The \underline{second best} results are underlined.}
\label{tab:performance_comparison on biocas 2022/2023}
\begin{tabular}{lccccccc}
\hline
\textbf{task} & \textbf{method} & \textbf{\#parameters} & \textbf{$S_p$ (\%)} & \textbf{$S_e$(\%)} & \textbf{$AS$ (\%)} & \textbf{$HS$(\%)} & $\textbf{TS} (\%)_{\pm{std}} $\\ 
\hline
\textbf{SPRSound  2022}  \\ 
\hline
\parbox[t]{2mm}{\multirow{7}{*}{\rotatebox[origin=c]{90}{Task1}}} 

 & CNN with MFCC \cite{babu2022multiclass} & 0.8M & 88.60 & 60.71 & 74.65 & 72.05 & 73.35 \\ 
 & Polymerized Feature Analysis \cite{zhang2022feature} & - & 83.39 & 66.10 & 74.75 & 73.75 & 74.25 \\ 
 & DenseNet169 with fine-tuning \cite{ma2022effective} & 14, 15 M & 72.92 & 76.60& 74.76 & 74.71 & 74.73 \\ 
 & ResNET-18 with STFT \cite{chen2022classify} & 11.5 M & \textbf{93.87} & 67.66 & 80.76 & 78.64 & 79.70 \\ 
 & ResNET-18 with Focal Loss \cite{li2022improving} & 12.2 M & 84.72 & \textbf{79.43} & 82.08 & 81.99& $82.03_{\pm{0.66}}$ \\ 
  & \textbf{CNN-TSA  (Ours)}  & 1.11 M & 90.39 & 75.46 & 82.93 & 82.26 & $82.59_{\pm{0.5}}$\\
 & \textbf{CNN-TSA +FBS(BS) $_{50\%freq}$(Ours)}  & 1.11 M & \underline{91.67} & 75.18 & \underline{83.42}&  82.61& $82.76 _{\pm{0.28}}$ \\
 & \textbf{CNN-TSA +FBS(BS) $_{75\%freq}$(Ours)}  & 1.11 M & 89.41 & 76.60 & 83.00&  82.51& $83.01_{\pm{0.43}}$\\
 & \textbf{CNN-TSA +FBS(IS)$_{50\%freq}$ (Ours)}  & 1.11 M &87.09  & \underline{79.15} & 83.12 &\underline{82.93} & $\underline{83.03} _{\pm{0.33}}$ \\ 
  & \textbf{CNN-TSA +FBS(IS)$_{75\%freq}$ (Ours)}  & 1.11 M &91.38  & 76.03 & \textbf{83.70} & \textbf{83.00} & $\textbf{83.35}_{\pm{0.3}}$ \\ 
\hline
\parbox[t]{2mm}{\multirow{7}{*}{\rotatebox[origin=c]{90}{Task2}}} 

 & Polymerized Feature Analysis \cite{zhang2022feature} & - & 79.86 & 81.41 & 81.99 & 81.74 & 81.97 \\ 
 & CNN with MFCC \cite{babu2022multiclass} & 0.8M & 83.56 & 83.55 & 83.56 & 83.56 & 83.56 \\ 
 & DenseNet169 with fine-tuning \cite{ma2022effective} & 14, 15 M & 88.25 & 81.70 & 84.99 & 84.85 & 84.91 \\ 
 & ResNET-18 with Focal Loss \cite{li2022improving} & 12.2 M & 84.72 & \textbf{93.19} & 88.96 & 88.76 & 88.86 \\ 
 & ResNET-18 with STFT \cite{chen2022classify} & 11.5 M & 89.58 & 88.94 & 89.26 & 89.26 & 89.26 \\ 
 & \textbf{CNN-TSA  (Ours)} & 1.11 M & 89.07 & 88.65 & 88.86 & 88.82 & $88.84_{\pm{0.55}}$ \\ 
 & \textbf{CNN-TSA +FBS$_{50\%freq}$(BS) (Ours)}  & 1.11 M & 86.81 & \underline{92.07} & 89.44 & 89.36 & $89.40_{\pm{0.37}}$ \\ 
 & \textbf{CNN-TSA +FBS$_{75\%freq}$(BS) (Ours)}  & 1.11 M & 88.37 & 90.50 & 89.43 & 89.42 & $89.43_{\pm{0.61}}$ \\ 
  & \textbf{CNN-TSA +FBS(IS)$_{50\%freq}$ (Ours)}  & 1.11 M & \underline{89.93} & 89.35 & \underline{89.64} & \underline{89.64}& $ \underline{89.64}_{\pm{0.35}}$\\
 & \textbf{CNN-TSA +FBS(IS)$_{75\%freq}$ (Ours)}  & 1.11 M & \textbf{91.06} & 88.06 & \textbf{89.86} & \textbf{89.84}&  $\textbf{89.85}_{\pm{0.54}}$\\

\hline
\textbf{SPRSound  2023} \\ 
\hline
\parbox[t]{2mm}{\multirow{5}{*}{\rotatebox[origin=c]{90}{Task1}}} 
& Dual-Input CNN \cite{pessoa2023pediatric} & 0.3M & 64 & 34 & 49 & 44 & 46.66 \\ 
 & ResNet (SCL)\cite{hu2023supervised} & 11.5M & - & - & - & - & 63.18 \\ 
 & Inception-residual \cite{ngo2023deep} & $>$12M & - & - & - & - & 66.66 \\ & \textbf{CNN-TSA  (Ours)} & 1.11 M & 91.10 & 53.53 & 72.31 & 67.41 & $69.86_{\pm{0.4}}$ \\ 
 & \textbf{CNN-TSA +FBS(BS)$_{50\%freq}$ (Ours)}  & 1.11 M & \textbf{93.78} & 51.95 &\textbf{72.86}  & 66.86 &$69.86_{\pm{0.41}}$  \\ 
 & \textbf{CNN-TSA +FBS(BS)$_{75\%freq}$ (Ours)}  & 1.11 M & \underline{91.46} & 53.75 &72.61  & 67.71 &$70.16 _{\pm{0.32}}$\\ 
 & \textbf{CNN-TSA +FBS(IS)$_{50\%freq}$ (Ours)}  &1.11 M & 83.93 & \textbf{59.76} & 71.84 &  \textbf{69.81}&  $\underline{70.83}_{\pm{0.29}}$\\
 & \textbf{CNN-TSA +FBS(IS)$_{75\%freq}$ (Ours)}  &1.11 M & 88.32 & \underline{57.36} & \underline{72.84} &  \underline{69.55}&  $\textbf{71.20}_{\pm{0.33}}$\\
\hline
\parbox[t]{2mm}{\multirow{5}{*}{\rotatebox[origin=c]{90}{Task2}}} 
& Dual-Input CNN \cite{pessoa2023pediatric} & 0.3M& 86 & 67 & 76 & 74 & 75.6 \\ 
 & ResNet (SCL) \cite{hu2023supervised} & 11.5M& - & - & - & - & 76.93 \\ 
 & Inception-residual \cite{ngo2023deep} &$>$12M & - & - & - & - & 80.97 \\ 
  & \textbf{CNN-TSA  (Ours)} & 1.11 M & 87.43 & 77.93 & 82.68 & 82.41 & $82.54_{\pm{0.44}}$ \\ 
 
 & \textbf{CNN-TSA +FBS(BS)$_{50\%freq}$ (Ours)}  & 1.11 M & 83.85 & \textbf{81.68} &  82.77& \textbf{82.75} & $82.76_{\pm{0.36}}$ \\ 
  & \textbf{CNN-TSA +FBS(BS)$_{50\%freq}$ (Ours)}  & 1.11 M & \underline{90.03} & 76.43 &  \underline{83.23}& 82.67 & $82.95_{\pm{0.27}}$ \\ 
 & \textbf{CNN-TSA +FBS(IS)$_{50\%freq}$ (Ours)}  & 1.11 M & 88.00 & \underline{78.08} & 83..04 & \underline{82.74} & $\underline{82.89}_{\pm{0.35}}$ \\
 & \textbf{CNN-TSA +FBS(IS)$_{75\%freq}$ (Ours)}  & 1.11 M & \textbf{90.94} & 76.92 & \textbf{83.93} & 82.47 & $\textbf{83.20}_{\pm{0.4}}$ \\
  \hline
\end{tabular}
\end{table*}

\section{Discussion}

\label{sec:discussion}
\begin{figure}[!b]
\centering
\includegraphics[width=1\linewidth]{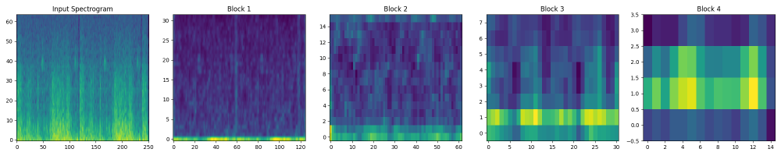}
\caption{Visualization of feature maps at different stages of the CNN backbone.}
\label{fig}
\end{figure}

This study demonstrates the effectiveness of a lightweight CNN architecture integrated with temporal self-attention and frequency band selection for high-performance respiratory sound classification.  Key contributions include the strategic placement of Temporal Self-Attention (TSA) and the implementation of importance-based Frequency Band Selection (FBS(IS)), each improving performance and efficiency.
The placement of TSA is critical to model performance (Table~\ref{tab:attn-placement}). As shown in Figure~\ref{fig}, early CNN layers extract low-level acoustic features, while deeper layers encode temporally structured representations. Applying TSA after the final convolutional block, especially following frequency aggregation, allows the model to focus on high-level, temporally aligned features, reducing sensitivity to low-level fluctuations. This placement improves both accuracy and efficiency by better modeling cyclic respiratory patterns with minimal added computational cost.

FBS(IS) complements TSA by enhancing both accuracy and computational efficiency. As demonstrated in Figure~\ref{fig:P}, selectively removing noisy or redundant frequency bands consistently improves classification performance across datasets. Peak results are achieved when retaining 50–75\% of frequency bands, indicating significant redundancy in the spectral input. FBS(IS) isolates the most discriminative bands while reducing computational demands by approximately 50\%, making it well-suited for real-time and embedded applications. The hyperparameter $\lambda$ further refines FBS(IS) by controlling the trade-off between maximizing overall relevance and minimizing class-wise variability (Figure~\ref{fig:lambda}). In binary classification, higher $\lambda$ values prioritize consistently informative bands, improving generalization. In contrast, multiclass tasks benefit from moderate $\lambda$ values that preserve inter-class variability. This flexibility enables task- and dataset-specific spectral adaptation. The effectiveness of TSA and FBS(IS) is further validated in ablation studies (Table~\ref{tab: ablation study on ICBHI}). TSA alone enhances temporal modeling with negligible GFLOPs overhead, while FBS(IS) improves accuracy and reduces computational cost by half. Compared to backward selection (FBS(BS)), FBS(IS) achieves similar or better performance without extensive evaluation, offering a more targeted and efficient frequency band selection strategy, and serving as a scalable alternative.

Experiments on age-specific models reveal reduced inter-patient variability and more structured acoustic patterns in pediatric cases, highlighting the value of demographically tailored diagnostic models to improve performance.

Comparative results (Tables~\ref{tab:performance_comparison on icbhi 2017} and \ref{tab:performance_comparison on biocas 2022/2023}) further affirm the strength of the proposed approach. Despite its small size (4.6M and 1.1M parameters), the model matches or exceeds large-scale transformer-based architectures like CLAP and AST (up to 190M parameters), demonstrating the value of spectro-temporal structure-aware modeling over scale alone, particularly in constrained settings.

Finally, integrating FBS(IS) into the transformer-based Patch-Mix CL model further improves performance (Table~\ref{tab:performance_comparison on icbhi 2017}), suggesting that frequency selection acts as input regularization. By simplifying spectral inputs and guiding attention toward relevant content, FBS(IS) enhances attention-based modeling of global time-frequency dependencies, highlighting its broader utility in respiratory sound analysis.

\section{Conclusion}
\label{sec:conclusion}
This study proposed a lightweight yet effective framework for automated respiratory sound classification, combining a CNN with temporal self-attention (CNN–TSA), an importance-guided Frequency Band Selection strategy (FBS(IS)), and age-specific models. The temporal self-attention module enhances the capture of relevant acoustic patterns, while FBS(IS) focuses computation on the most informative spectral bands, achieving higher accuracy and faster convergence compared to backward selection, along with a 50\% reduction in computational cost. The method is model-agnostic, demonstrating improvements even when applied to transformer architectures. Additionally, age-adaptive training on the ICBHI dataset mitigates demographic variability. Together, these components yield strong, generalizable performance across tasks and datasets, making the framework well-suited for real-time clinical deployment. Future work will investigate dynamic estimation of band importance and extend FBS(IS) to alternative time–frequency representations.

\section*{References}
\bibliographystyle{IEEEtran}
\bibliography{TBE}

\end{document}